\documentclass[11pt]{article}
\usepackage{amsfonts}
\usepackage{bbm}
\usepackage{graphicx}
\newcommand {\beq}{\begin{equation}}
\newcommand {\eeq}{\end{equation}}
\newcommand {\beqa}{\begin{eqnarray}}
\newcommand {\eeqa}{\end{eqnarray}}

\begin{document}

\title{Entanglement may enhance the channel capacity in arbitrary dimensions}

\author{E.~Karpov, D.~Daems and N.~J.~Cerf\\
\small{\em{Quantum Information and Communication, Ecole
Polytechnique}}, \\
\small{\em{CP 165, Universit\'e Libre de Bruxelles, 1050
Brussels, Belgium}}}
\date{}
\maketitle

\begin{abstract}
We consider explicitly two examples  of $d$-dimensional quantum channels with
correlated noise and show that, in agreement with previous results
on Pauli qubit channels,
there are situations where maximally entangled input states achieve
higher values of the output mutual information than product states.
We obtain a strong dependence of this effect on the nature of
the noise correlations as well as on the parity of the space dimension,
and conjecture that when entanglement
gives an advantage in terms of mutual information,
maximally entangled states achieve the channel capacity.
\end{abstract}

\section{Introduction}


The
evaluation of the amount of \emph{classical information} which can
be reliably transmitted by \emph{quantum states} is a major problem
of quantum information. Early works in this direction, devoted
mainly to memoryless channels, for which consecutive signal
transmissions through the channel are not correlated,
allowed the determination of capacities in many instances
\cite{H73}-
\cite{MPZ01}
and  proved their additivity.
Recently much attention was given to quantum channels
with memory \cite{DWM03}-
\cite{G05}
in the hope that the capacity of such channels be superadditive
and  could  be enhanced  by using entangled
input states. Contrary to this expectation,
for bosonic memory channels without input energy constraints,
the additivity conjecture was proven leaving no hope
to enhance the channel capacity using entangled inputs \cite{GM05}.
However, for more realistic situation where the energy of input Gaussian states
is finite, it was shown that entangling two consecutive uses of the channel
with memory introduced by a correlated noise enhances the
overall channel capacity \cite{CCMR05,RSGM05}.
For each value of the noise correlation parameter,
there exists an optimal degree of entanglement
(not necessarily maximal entanglement)
that maximizes the channel capacity.
Other examples of quantum channels with memory introduced by a correlated noise
include qubit Pauli channels \cite{Mac02,Mac04}.
For these channels it was shown that if the noise correlations
are stronger than some critical value,
maximally entangled input states enhance the channel capacity compared
to product input states.
Quantum channels with correlated noise in
dimensions $d>2$ were not considered in the literature in this
context, except for \cite{KM06}, which appeared after our work
was completed and presented, where a class of $d$-dimensional
quantum channels with memory is considered for which maximally entangled states
maximize the channel capacity beyond some memory threshold.
The $d$-dimensional channels correspond to a kind of intermediate system
between the qubit and the Gaussian channels.
Therefore, we expect to find new features
that this intermediate dimensionality can add to the known facts.
We shall consider $d$-dimensional quantum channel which are
generalizations of the Pauli qubit channels studied in \cite{Mac02,Mac04}.
We start with the introduction of the classical capacity of quantum channels,
consider explicitly two examples of $d$-dimensional quantum channels
with memory and present results on their capacity.

\section{Capacity of quantum channels with correlated noise}

The action of a transmission channel on an initial quantum state
described by density operator $\rho$ is given by a linear
completely positive (CP) map
${\mathcal E}:  \rho \rightarrow {\mathcal E}(\rho)$.
The amount of classical information which can be reliably transmitted
through a quantum channel is given by the Holevo-Schumacher-Westmoreland
bound \cite{H73,SW97} as the maximum of mutual information
\begin{equation}\label{HSW}
   \chi(\mathcal E)
        = \max_{\{P_i,\rho_i\}} I(\mathcal E)
\end{equation}
taken over all possible ensembles $\left\{P_i,\rho_i\right\}$
of input states $\rho_i$ with {\it a priori} probabilities
$P_i>0, \quad \sum_i P_i =1$.
The mutual information of an ensemble $\left\{P_i,\rho_i\right\}$
is defined as
\begin{equation}\label{I}
   I(\mathcal E(\{P_i,\rho_i\}))
     =  \left[ S\left(\sum_i P_i\mathcal E(\rho_i)\right)
                     -\sum_i P_i S(\mathcal E(\rho_i))
                \right]
\end{equation} where $S(\rho) = - {\rm Tr} [\rho \log_2\rho]$ is
the von Neumann entropy.
If we find a state $\rho_*$ which minimizes the output entropy
$S({\mathcal E(\rho_*)})$ and replace the first term in (\ref{I})
by the largest possible entropy given by the entropy of the
maximally mixed state, we obtain the following bound
\begin{equation}\label{bound}
  \chi({\mathcal E})\le \log_2(d)-S({\mathcal E}(\rho_*))
\end{equation}
where $d^2$ is the dimension of $\rho$.
Generalizing the arguments of \cite{Mac04}, this bound can be shown \cite{KDC06} to
become tight for the channels which we consider.

In general, being a linear CP map,
any quantum channel can be represented by an operator-sum:
${\mathcal E}(\rho)
= \sum_k A_k\rho A_k^\dag$, $\sum_k A_k^\dag A_k= \mathbbm{1}$.
In order to describe $n$ uses of the same quantum channel,
we have to consider the Hilbert space of the initial states,
which is a tensor product such that
$\rho \in \mathcal H^ {\otimes n}$.
The repeated use ($n$ times) of the channel is CP map ${\mathcal E}_n$
\begin{equation}
  {\mathcal E}(\rho) = \sum_{k_1,\dots k_n} A_{k_1\dots k_n}\rho A_{k_1\dots k_n}^\dag
\end{equation}
represented by the Kraus operators acting in $d\times n$-dimensional space
The channel is memoryless if the Kraus operators can be factorized according to
\begin{equation}\label{product}
  {\mathcal E}_n(\rho)
    = \sum_{k_1\dots k_n} p_{k_1\dots k_n}(A_{k_1}
               \otimes\dots \otimes A_{k_n})\rho
               (A_{k_1}^\dag\otimes\dots \otimes A_{k_n}^\dag) .
\end{equation}
where each operator $A_k$ represents one single use of the channel
and the normalized probability distribution $p_{k_1\dots k_n}$ is
  factorized  into
probabilities which are independent for each use of the channel.
On the other hand, a memory effect is introduced  when
correlations are present  between consecutive uses of the channel,
e.g., when each use of the channel depends on the
preceding one in such a way that $p_{k_1\dots k_n}$ is given by a product
of conditional probabilities
$  p_{k_1\dots k_n}=p_{k_1}p_{k_2|k_1}\dots p_{k_n|k_{n-1}}$.
Indeed, the correlations between the consecutive uses of the
channel act as if the channel ``remembers'' the first signal and
acts on the second one using this ``knowledge''. This type of
channels is called a Markov channel as the probability $  p_{k_1\dots k_n}$
corresponds to a Markov chain of order 2.

Following \cite{Mac02, Mac04} one can introduce
a Markov type of memory effect by choosing
probabilities $ p_{ij}$ which include a correlated noise:
  $p_{ij}=(1-\mu)\,p_ip_j+\mu\, p_i\delta_{ij}$.
The memory parameter $\mu\in[0,1]$ characterizes the correlation
``strength''. Indeed, for $\mu = 0$, the probabilities of two
subsequent uses of the channel are independent, whereas for $\mu=1$, the
correlations are the strongest ones.
We shall consider channels given by a product of pairwise correlated channels
${\mathcal E}_{2n}= {\mathcal E}_2^{\otimes n}$
where ${\mathcal E}_2$ is determined by (\ref{product}).
These channels are not strictly Markov:
due to the pair wise correlations, $2n$ quantum states
sent are split into $n$ consecutive pairs.
The actions of the channel on the states belonging to the same pair
are correlated according to $p_{ij}$,
whereas, the actions of the channel on
the states from different pairs are uncorrelated.
However, even these ``limited'' (within each pair) correlations
result in the advantages of using entangled input states
as it was shown in \cite{Mac02,Mac04}.

\section{Model and results}

We shall study $d$-dimensional Heisenberg channels \cite{Cerf00}
which may be considered as a generalization Pauli qubit channels.
The Kraus operators are therefore given by the ``error''
or ``displacement'' operators acting
on $d$-dimensional states.
  \begin{equation}\label{U}
    U_{m,n}
       = \sum_{k=0}^{d-1}e^{\frac{2\pi i}{d} kn}\,|k+m\rangle\langle k|,
  \end{equation}
where the index $m$ characterizes the displacement (of
a mode) or cyclic shift of the basis vectors of the pointer basis
by analogy with the bit-flip,
and the index $n$ characterizes the phase shift.
The displacement operators
form a Heisenberg group \cite{F95} with commutation relation
\begin{equation}\label{cr}
 U_{m,n}U_{m',n'}
    = e^{2\pi i (m'n-mn')/d}U_{m',n'}U_{m,n}.
\end{equation}
For two uses of a $d$-dimensional Heisenberg channel the CP map
(\ref{product}) becomes
  \begin{equation}\label{emn}
    {\mathcal E}_2(\rho)
      = \sum_{m,n,m',n'=0}^{d-1}p_{m,n,m',n'}
     \times (U_{m,n}\otimes U_{m',n'})\,\rho\,
         (U_{m,n}^\dag\otimes U_{m',n'}^\dag),
  \end{equation}
where $\rho$ is a $d^2 \times d^2$ density matrix.
The Markov-type joint probability reads
 \begin{equation}\label{pmn}
    p_{m,n,m',n'}
       =  (1-\mu)q_{m,n}\,q_{m',n'}
       +  \mu\,q_{m,n}\,\delta_{m,m'}\,
       ((1-\nu)\,\delta_{n,n'}+\nu\,\delta_{n,-n'}).
  \end{equation}
Notice the presence of a product of two Kronecker' deltas
representing the noise correlations
separately for displacements (index $m$) and for phase shifts (index $n$).
In addition, we introduce both, phase correlations ($\delta_{n,n'}$)
as well as phase anticorrelations ($\delta_{n,-n'}$)
with a new parameter $\nu$ characterizing the type of
the phase correlations in the channel.
For $d=2$ such a distinction disappears
as phase correlations $\delta_{n,n'}$
and phase anticorrelations $\delta_{n,-n'}$ coincide.
Note that for infinite dimensional bosonic Gaussian channel
the phase anticorrelations provided an enhancement of the channel capacity
by entangled states \cite{CCMR05}.
We consider two types of $d$-dimensional
channels, characterized by the following sets of $q_{m,n}$:

\vskip 0.5 cm
\emph{A. Quantum depolarizing (QD) channel}:
 $q_{m,n} = p $ if $m=n=0$ and  $q_{m,n}=q$ otherwise.
As  $q= (1-p)/(d^2-1)$ the channel is characterized by a single parameter
$\eta=p-q \in [-1/(d^2-1),1]$ reminiscent of the``shrinking factor''
of the two-qubit QD channel.

\vskip 0.5cm
\emph{B. Quasi-classical depolarizing (QCD) channel}:
$q_{m,n}= q_m $ with $q_m= p$ if $m=0$ and $q_m=q$ otherwise.
The probabilities of the displacements of the same mode $m$ are equal
regardless of the phase shift (determined by $n$) and the probability of
``zero'' displacement ($m=0$)
differs from the others that are  equal.
We call this channel quasi-classical as
a classical depolarizing channel changes the amplitude of the modes
of a classical signal with some probability,
but there is no quantum phase in classical signals.
Since $q  =(1-dp)/(d(d-1))$ the channel is characterized
by the parameter $\eta=d(p-q) \in  [-1/(d-1),1]$.

For these  $d$-dimensional channels,  following \cite{Mac04}
and using their covariance
we have proven \cite{KDC06} that the bound (\ref{bound}) is tight, i.e.,
 in order to determine the capacity
of the channels we have to find an \emph{optimal} state $\rho_*$
that minimizes the output entropy $S(\mathcal E_2(\rho_*))$.

By analogy with the two-dimensional case \cite{Mac02,Mac04} we look
for an optimal $\rho_*$ as a pure input state
$\rho_{\rm in}=|\psi_0\rangle\langle \psi_0|$ where
\begin{equation}\label{iniphi}
  |\psi_0\rangle = \sum_{j=0}^{d-1}\alpha_j e^{i\phi_j}|j\rangle|j\rangle,
                   \quad \alpha_j \ge 0, \quad
   \sum_{j=0}^{d-1}\alpha_j^2 =1.
\end{equation}
This  ansatz allows us to go from a product state to a maximally entangled state
by changing the parameters $\alpha_j$ and $\phi_j$.
Indeed, the choice $\alpha_j=\delta_{j,0}$ and $\phi_j=0$
results in a product state whereas the choice
$\alpha_j=1/\sqrt{d}$ and $\phi_j=0$ results in a maximally entangled state.
Taking into account the form (\ref{U}) of the displacement operators $U_{m,n}$,
the probability distribution $p_{m,n,m',n'}$ (\ref{pmn})
and the probability parameters $q_{m,n}$ for both channels,
we evaluate the action of the channel given by Eq. (\ref{emn})
on the initial state $|\psi_0\rangle\langle\psi_0|$ in the form (\ref{iniphi}).
Then 
we diagonalize the output states
and find their von Neumann entropy that allows us to obtain the mutual information
according to Eq. (\ref{I}).
We evaluate the action of the QD channel given by
Eqs. (\ref{emn}-\ref{pmn}) and of the QCD channel given by
Eqs. (\ref{emn}-\ref{pmn})
on a pure initial state given by Eq. (\ref{iniphi}).
The analytic results of these evaluations are presented elsewhere \cite{KDC06}.
Here we shall present and discuss some figures displaying these analytic
results, but before we discuss whether these results provide
an \emph{optimal} $\rho_*$.

\begin{figure}[ht]
\begin{center}
  \includegraphics[width=0.47\textwidth]{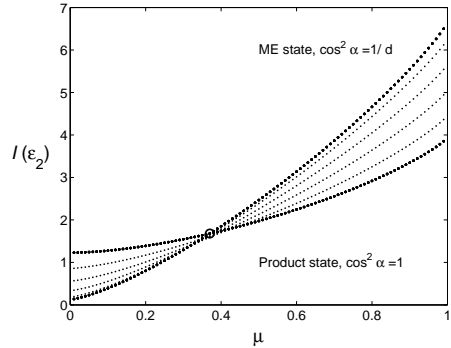}
  \caption{Mutual information $I({\mathcal E}_2(\rho_\alpha))$
           as function of the memory  parameter
           $\mu$ for QCD channel with $\eta=0.4$ for different
           values of the optimization parameter $\alpha$.
          }

\end{center}
\end{figure}

The task of finding an \emph{optimal} $\rho_*$
becomes easier for the quasi-classical depolarizing channel
because we can restrict our search from the whole space to a certain subclass.
In order to show this we note, following \cite{Mac04},
that the averaging operation ${\mathcal F}$
\begin{equation}
{\mathcal F}(\rho)=\frac{1}{d}\sum_{n=0}^{d-1}
          (U_{0,n}\otimes U_{0,n})\rho(U^\dag_{0,n}\otimes U^\dag_{0,n})
\end{equation}
does not affect the QCD-channel in the sense that
$\mathcal E_2\circ{\mathcal F}=\mathcal E_2$.
Then, if $\rho_*$ is an \emph{optimal} state
then ${\mathcal F}(\rho_*)$ is also an optimal state.
Therefore we can restrict our search from the whole space
${\mathcal H}^{\otimes 2}$
to ${\mathcal F}({\mathcal H}^{\otimes 2})$.
Finally, using (\ref{U}), it is straightforward to show that
any state  from ${\mathcal F}({\mathcal H}^{\otimes 2})$
is a convex combination of pure states $|\psi_m\rangle\langle\psi_m|$ where
\begin{equation}\label{state}
       |\psi_m\rangle
        =\sum_{j=0}^{d-1} \alpha_j e^{i\phi_j}|j\rangle |j+m\rangle,
          \quad \alpha_j\in {\mathbb R}, \quad \sum^{d-1}_{j=0} \alpha_j^2=1 .
\end{equation}
Restricting our search to the states of the form (\ref{state})
we reduce the number of real optimization parameters from $(2d)^2$ to $2d$,
which can still be a large number.
In order to reduce this number to 1,
we consider the following ansatz
\begin{equation}\label{ansatz}
  |\psi(\alpha)\rangle
         =\cos\alpha|00\rangle + \frac{\sin\alpha}{\sqrt{d-1}}
         \sum^{d-1}_{j=1}|j\, j\rangle ,
  \end{equation}
interpolating between the product state ($\cos\alpha=0$)
and the maximally entangled state ($\cos^2\alpha=1/d$).
Using the one-parameter family of input states
$\rho_\alpha=|\psi(\alpha)\rangle\langle\psi(\alpha)|$,
in Fig.~1 we present the mutual information $I(\mathcal E_2(\rho_\alpha))$
for different values of $\alpha$.
The mutual information is monotonously modified when $\alpha$ goes
from a product state to a maximally entangled state,
whereas the crossover point $\mu_c$  stays intact.
However, we cannot guarantee that
no other entangled state
minimizes the entropy $S(\mathcal E_2(\rho))$ and provides therefore
the maximum of the mutual information.

In the sequel, we present numerical graphs based on our analytic results
obtained for product states and maximally entangled states
as candidates for the \emph{optimal} $\rho_*$.
The mutual information $I(\mathcal E_2)$ is depicted in Fig. 2 as a function
of the memory parameter $\mu$ for both these states
for QD channel for various dimensions.
The curves in Fig.~2 (a) correspond to the strongest phase anti-correlations
expressed by $\delta_{n,-n'}$ in (\ref{pmn}) and correlation parameter $\nu=1$.
The curves in Fig.~2 (b) correspond to the strongest phase correlations expressed by
$\delta_{n,n'}$ in (\ref{pmn}) and the correlation parameter $\nu=0$.
\begin{figure}[h]
\begin{center}
\flushleft  (a) \hskip 9cm  (b)

\includegraphics[width=0.45\textwidth]{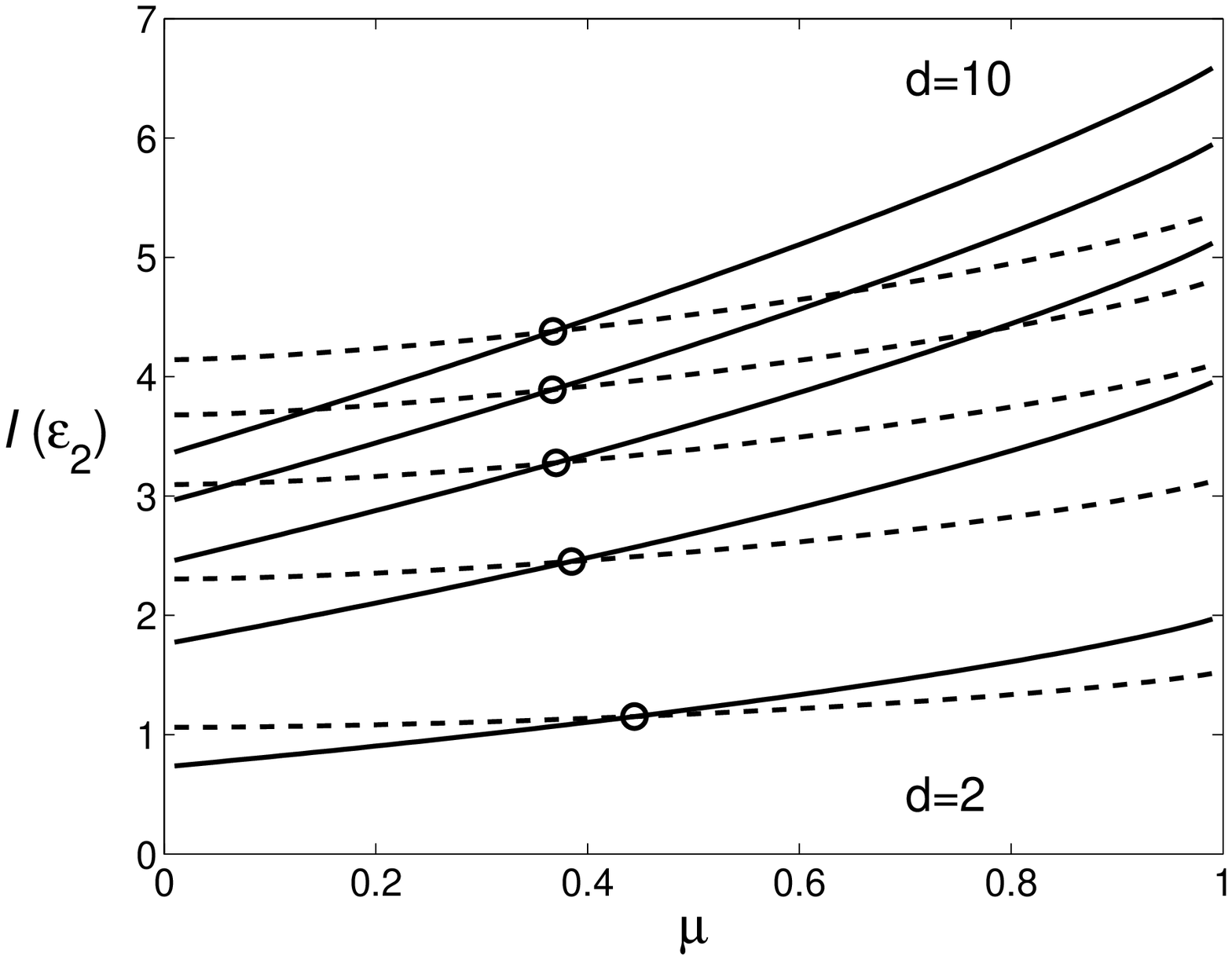}
\hskip 0.5 cm
 \includegraphics[width=0.45\textwidth]{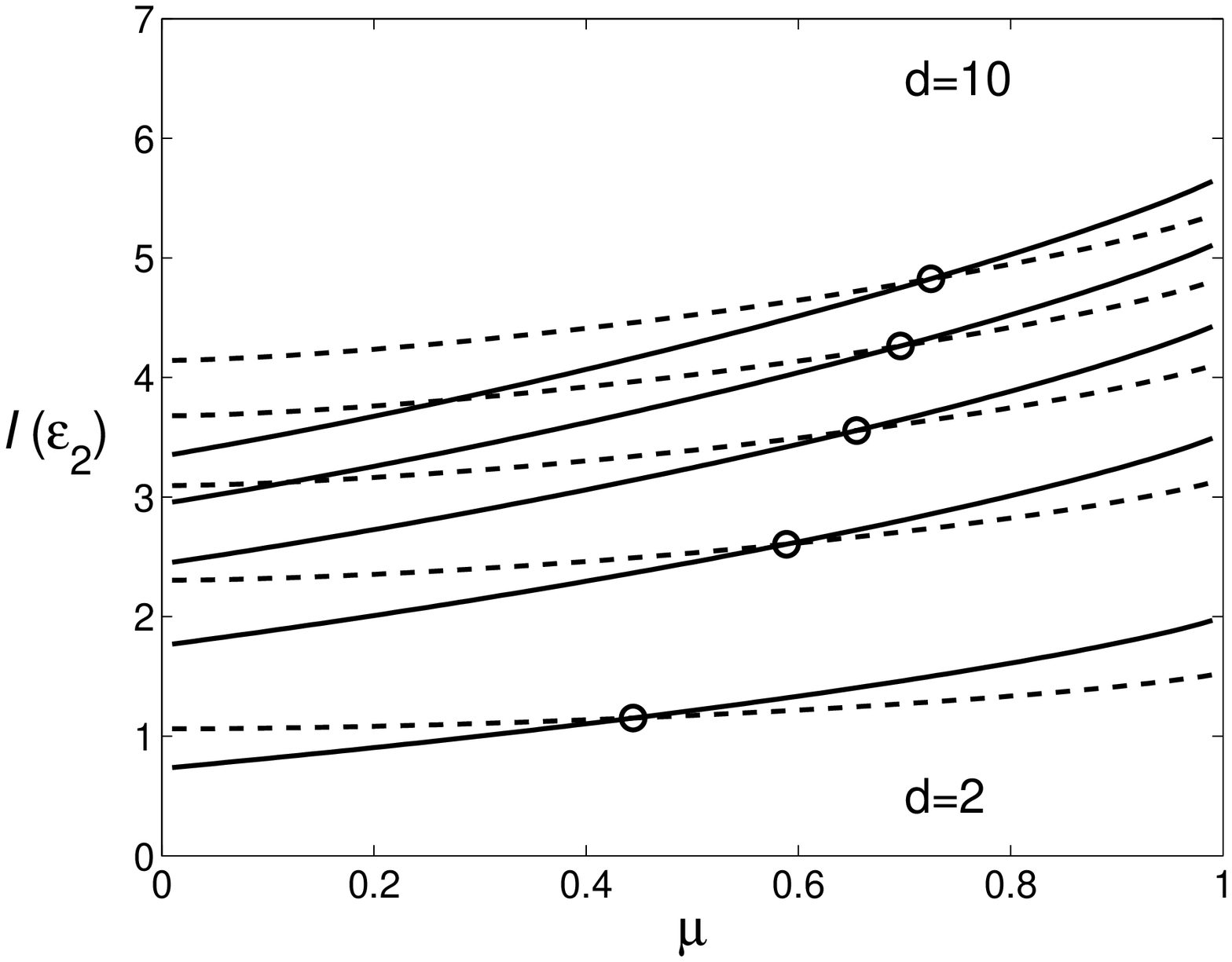}

\flushleft  (c) \hskip 9cm  (d)

\includegraphics[width=0.45\textwidth]{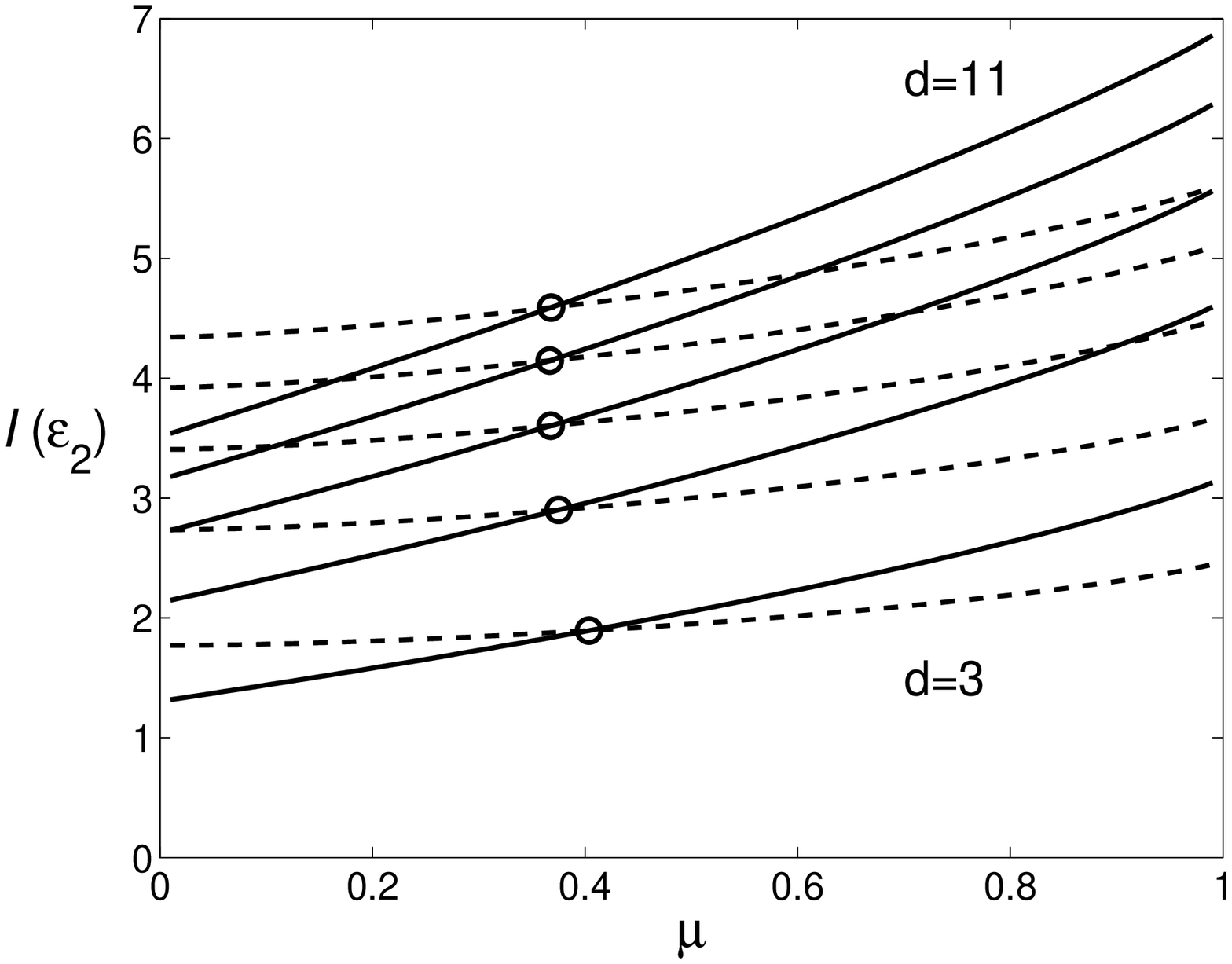}
\hskip 0.5 cm
(d) \includegraphics[width=0.45\textwidth]{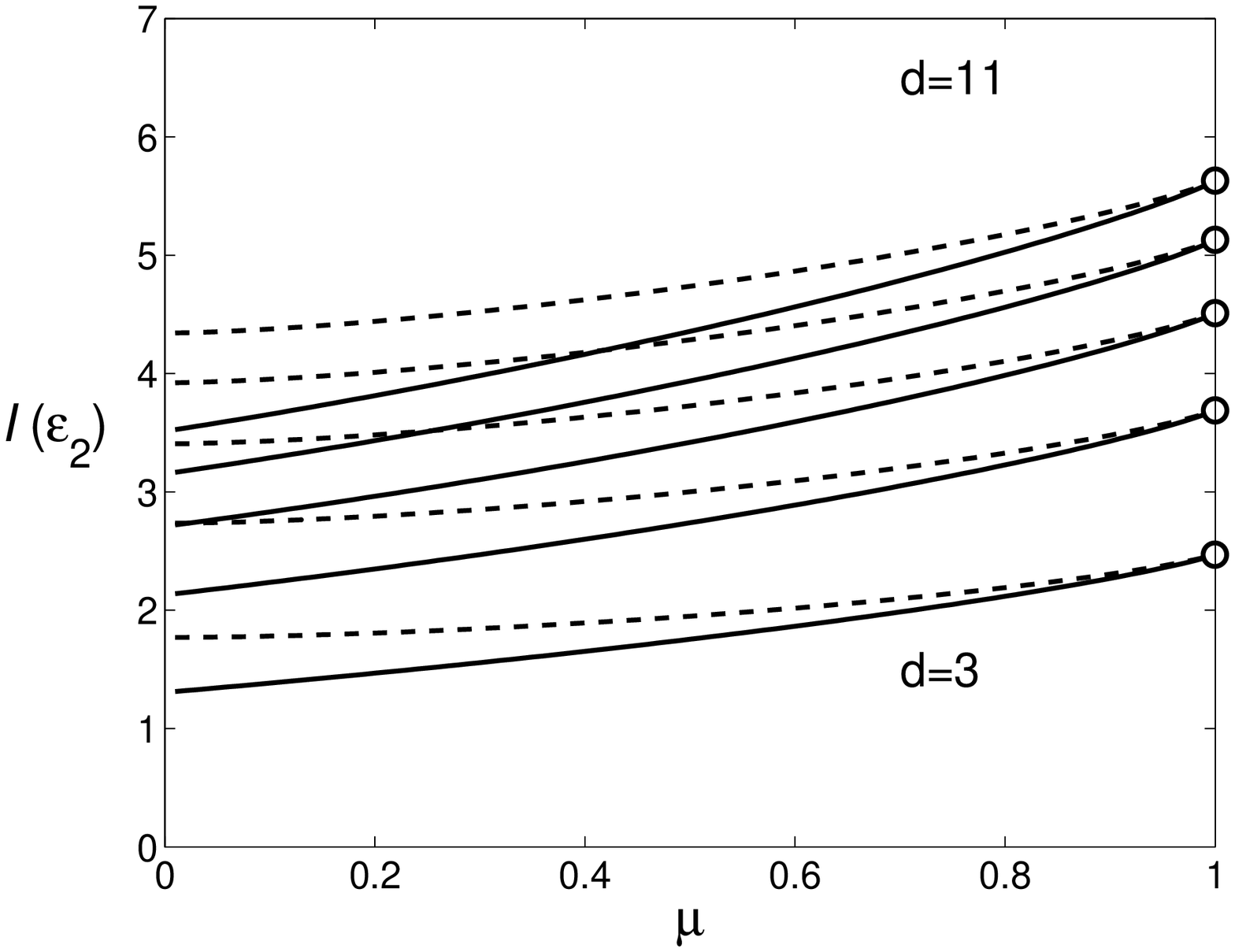}
\caption{Mutual information $I({\mathcal E_2}(\rho))$ as function of the memory
         parameter $\mu$ for QD ($\eta=0.8$) entanglement-friendly (a, c) with $\nu=1$
         and entanglement-non-friendly (b, d) with $\nu=0$ channels
         for different \emph{even} (a, b) and \emph{odd} (c, d)
         dimensions $d$.}
\end{center}
\end{figure}
In both figures for all drawn dimensions we see a crossover point,
the ``$\mu$'' coordinate of which we denote $\mu_c$.
For $\mu\in [0,\mu_c[$ the product states provide higher value of mutual entropy
and for $\mu\in ]\mu_c,1]$ the maximally entangled states do. In addition,
in Fig.~2 (a) we observe that with an increasing dimension
the crossover points move toward smaller $\mu$ thus
widening the interval where maximally entangled states provide
higher values of the mutual information than product states do.
For this reason we call the $\nu=1$ version of the channel
``entanglement-friendly''.
An opposite effect can be seen in Fig.~2 (b) where the crossover point
moves toward higher values of $\mu$ with the increasing dimension
of the space of states
thus shrinking the interval where entangled states perform better.
The $\nu=0$ version of the channel is  thus called ``entanglement-non-friendly''.
We note that for $d=2$ this difference between the two types of channels
disappears and we recover the result obtained in \cite{Mac02}.

In order to see the effect of the phase correlations for higher dimensions
we draw in Fig.~3 (a) the $\mu$ coordinate of the crossover point, $\mu_c$,
as a function of $d$.
We observe that only strongly anticorrelated phases ($\nu \approx 1$)
provide ``entanglement-friendly'' channel so that with increasing dimension
the interval of $\mu$'s that are favorable for entangled states increases.
In addition, even for $\nu=1$ this increase continues only up to certain $d$,
after which the interval of $\mu$ begins to shrink with increasing $d$.

We note that $\mu_c$ depends also on the ``shrinking factor'' $\eta$.
With increasing $\eta$ the slope of the curves, which are drawn in Fig.~3 (a)
for $\eta=0.8$ would become steeper and the upper curves,
corresponding to the small values of the phase  correlation parameter $\nu$
would cross the level $\mu_c=1$ at some $d<100$.
Hence for higher dimensions there is no values of $\mu$ for which entangled
input states may have any advantage at all.

For the QCD channel the result is  similar, hence  we present it only on
Fig.~3~(c) which shows that for the ``entanglement-non-friendly'' version
($\nu=0$) the advantages of entangled states completely disappear in higher
dimensions.

As the result, we conclude that for even dimensions,
the advantages of entangled states
are more essential for low (but not always lowest) dimensions,
anticorrelated phases, smaller values of $p$, and QD channel.

\begin{figure}[h]
\begin{center}
\flushleft  (a) \hskip 9cm  (b)

\includegraphics[width=0.45\textwidth]{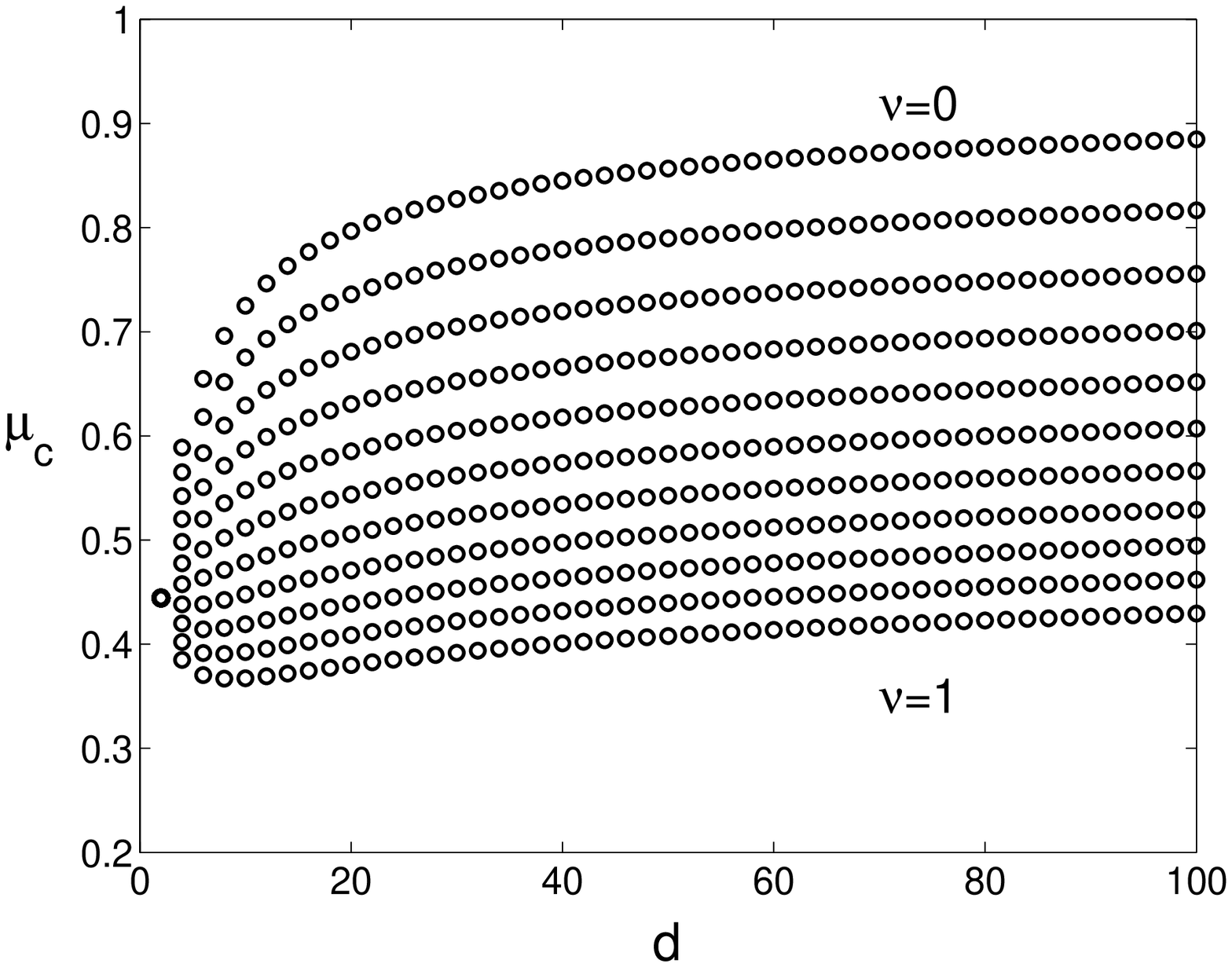}
\hskip 0.5 cm
\includegraphics[width=0.45\textwidth]{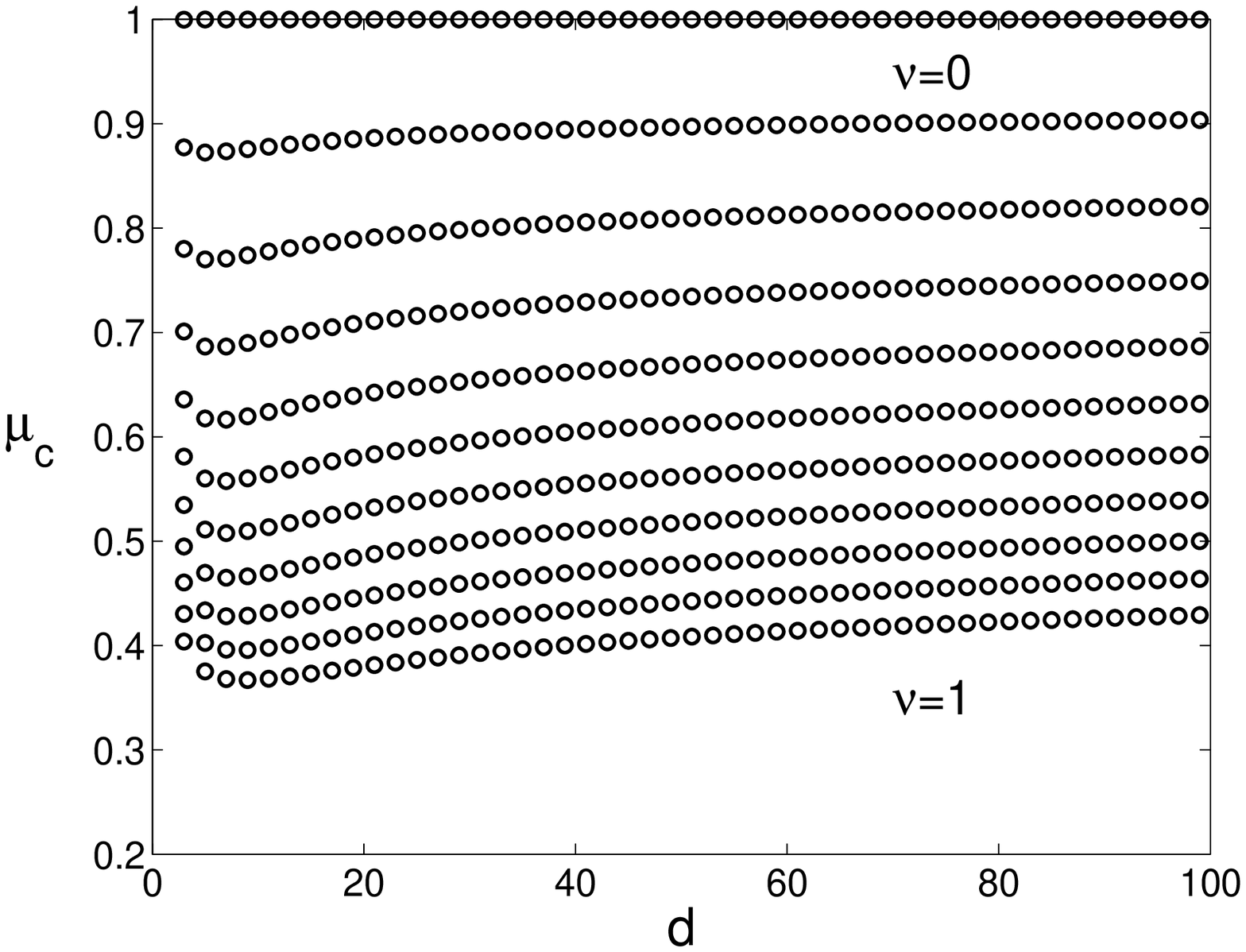}

\flushleft  (c) \hskip 9cm  (d)
\includegraphics[width=0.45\textwidth]{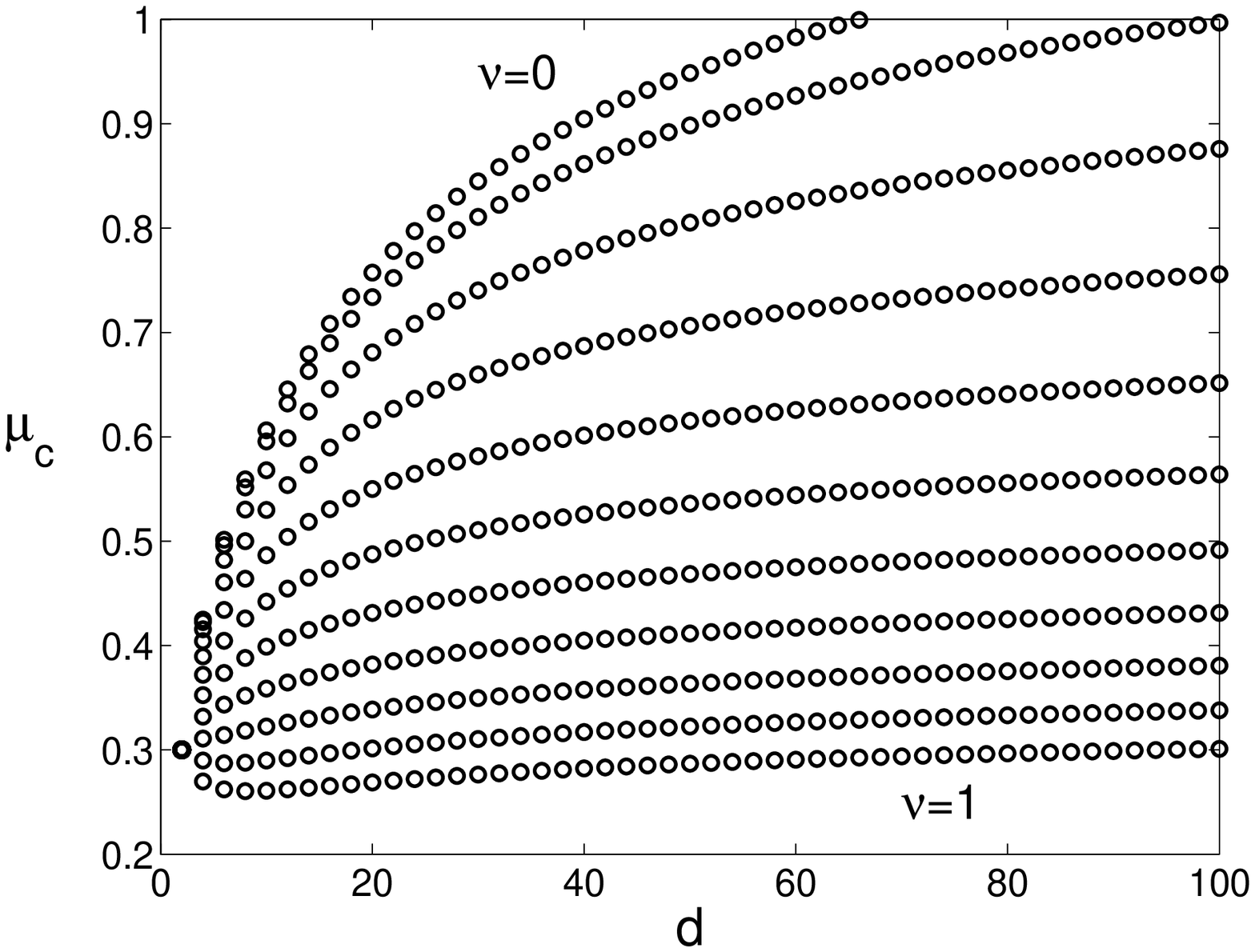}
\hskip 0.5 cm
\includegraphics[width=0.45\textwidth]{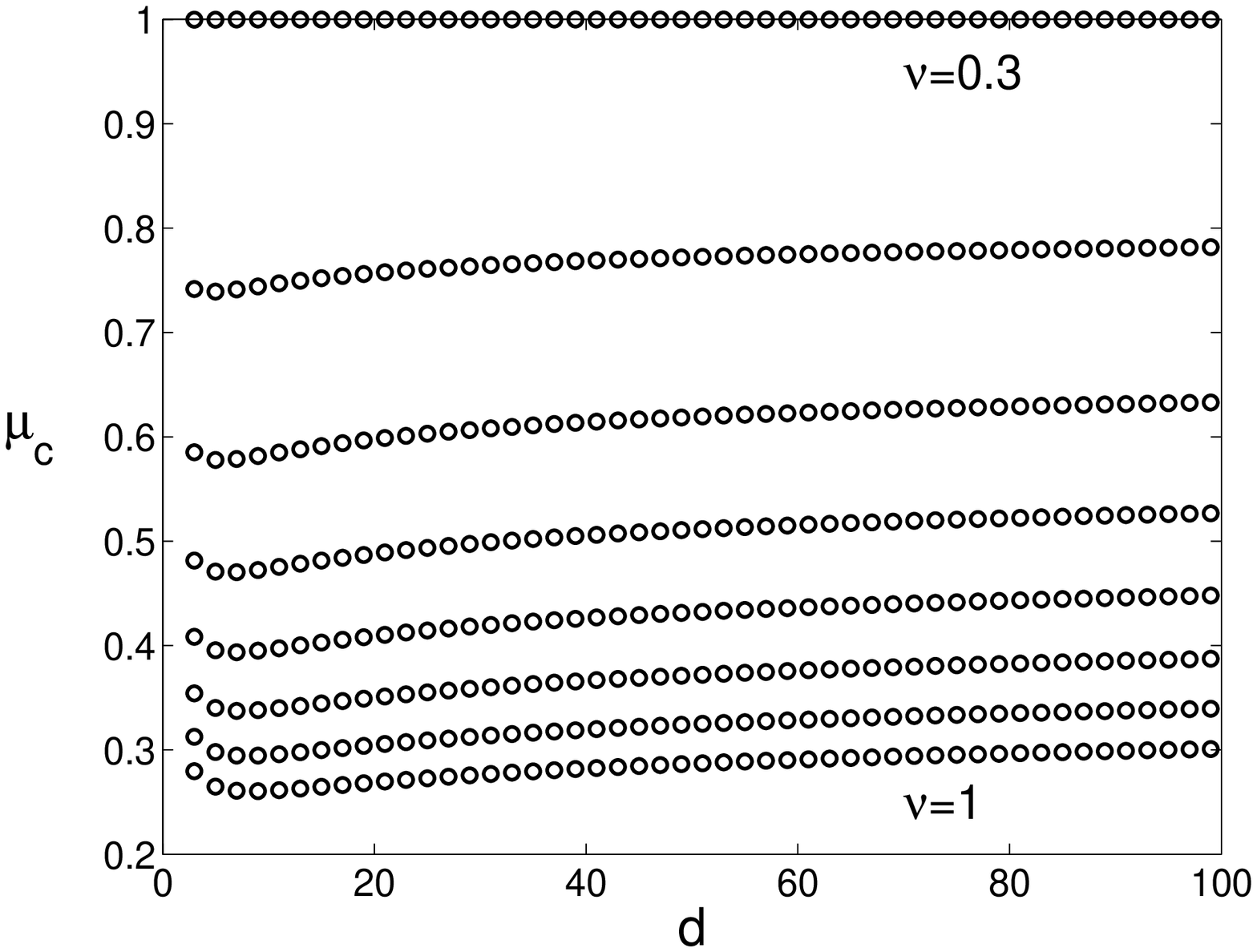}
\caption{Crossover point $\mu_c$ vs. \emph{even} (a,c)
         and \emph{odd} (b,d)
         space dimension for QD  channel (a,b) with  $\eta=0.8$
         and QCD channel (c,d) with $\eta=0.3$
         for different phase correlations $\nu$.
         }
\end{center}
\end{figure}

Our analytic formulas for \emph{even} and \emph{odd} dimensions are different.
However due to the factor $(1-\nu)$ the case of entanglement-friendly
($\nu=1$) version of both QD and QCD channels for \emph{odd} dimensions
differs from Fig.~2 (a) only in the position of the curves
whereas qualitatively the pictures are the same as displayed in Fig. 2 (c).
In the entanglement-non-friendly version ($\nu=0$) of the QD channel
in \emph{odd} dimensions shown in Figs.~2 (d) and 3 (b)
the crossover points  lay on the line $\mu=1$.
Therefore, effectively there is no crossover as $\mu$ cannot be larger than 1.
In this case for all $\mu$ the maximally entangled input states do not provide
higher values of the mutual information than the product states do.
For the QCD \emph{odd}-dimensional channel the picture is similar however,
the upper horizontal line $\mu = 1$ is achieved even for a non vanishing value
of the ``friendness'' parameter, $\nu =0.3$.

\section{Conclusion}

We have considered two examples of $d$-dimensional quantum channels
with a memory effect modeled by a correlated noise.
We have shown the existence of the crossover points
separating the intervals of the memory parameter $\mu$
where ensembles of maximally entangled input states
or product input states provide
higher values of the mutual information.
This result is the same as in the 2-dimensional case.
However it always holds only for channels
(which we call ``entanglement-friendly'' channels)
with a particular kind of phase correlations, namely anticorrelations.
For these channels the crossover point moves with increasing $d$
towards lower values of the memory parameter
thus widening the range of correlations
where maximally entangled input states enhance the mutual information.
For usual phase correlations the situation is opposite, namely,
for higher dimensions of the space the crossover point is shifted towards
$\mu_c=1$ so that only for higher degrees of correlations maximally
entangled input states have advantages.
In addition, for these ``entanglement-non-friendly'' channels
the crossover point completely disappears for higher dimensions
so that product input states
always provide higher values of the mutual information
than maximally entangled input states.
Therefore we conclude that the type of phase correlations  strongly affects
this entanglement assisted enhancement of the channel capacity.

We have observed that the parity of the dimension of the space
of initial states makes an important difference in the
``entanglement-non-friendly'' channels.
Not only the curves of the mutual information vs. the memory parameter
for odd dimensions are shifted with respect to the curves for even dimensions,
but also for $\nu=0$ in all odd dimensions maximally entangled input states
are always worse than product states.
Strikingly, the channels with anticorrelated noise do not feel the parity of the
space at all (in Fig. 3, compare the curves $\nu=1$ from (a) with (b)
and from (c) with (d)). However, any non vanishing degree of
the ``entanglement-non-friendly'' correlations
reveals the parity effect.

The anticorrelated phases remind us
the bosonic Gaussian channels considered in \cite{CCMR05}
where the $p$ quadratures are correlated while the $q$ quadratures are anticorrelated.
However, the existence of the crossover point is a significant
difference with the case of the Gaussian channels
for which each value of the noise correlation parameter
determines an optimal degree of entanglement (different from maximal entanglement)
maximizing the mutual information. A challenging open problem is to find
a link between these results for $d$-dimensional channels and the results obtained in
\cite{GM05,CCMR05,RSGM05} for Gaussian channels
with finite energy input signals.

Although we have shown that for certain cases of $d$-dimensional channels
maximally entangled states provide higher values of mutual information than
product states, a full proof of the optimality of maximally entangled input states
is still missing. However if it is true, the presented parametrization illustrating
a ``monotonous'' deformation of the curves of mutual information vs. the memory parameter
during the transition from product states to maximally entangled states shows
that at $\mu_c$ the optimal state ``jumps'' from the product to the maximally entangled state.
The ``sharp'' character of this transition is due to the fact that
the crossover points stay intact during the deformation of the curves.

The authors acknowledge the supported of the European Union projects SECOCQ
              (grant No. IST-2001-37559), CHIC (grant No. IST-2001-33578),
               and  QAP (contract 015848).

\end {document}